%% file: hpc_power_analysis.tex
\documentclass[letterpaper,twocolumn,10pt]{article}
\usepackage[pdftex]{graphicx}
\usepackage{usenix,epsfig,url}

\begin{document}

\title{Catch Me If You Can: Using Power Analysis to Identify HPC Activity}
\author{Bogdan Copos \\ 
Lawrence Berkeley National Laboratory\\
and University of California, Davis\thanks{Though this manuscript is only being published in 2020, because the authors didn't get around to doing so earlier, this work was actually done in 2015--2017 while Bogdan Cops was a graduate researcher at University of California, Davis and Lawrence Berkeley National Laboratory}\\ 
bcopos@ucdavis.edu
\and Sean Peisert\\
Lawrence Berkeley National Laboratory\\
and University of California, Davis\\
sppeisert@lbl.gov}





\maketitle

\subsection*{Abstract}
Monitoring users on large computing platforms such as high performance computing (HPC) and cloud computing systems is non-trivial.
Utilities such as process viewers provide limited insight into what users are running, due to granularity limitation, and other sources of data, such as system call tracing, can impose significant operational overhead.
However, despite technical and procedural measures, instances of users abusing valuable HPC resources for personal gains have been documented in the past~\cite{hpcbitmine}, and systems that are open to large numbers of loosely-verified users from around the world are at risk of abuse.
In this paper, we show how electrical power consumption data from an HPC platform can be used to identify what programs are executed.
The intuition is that during execution, programs exhibit various patterns of CPU and memory activity.
These patterns are reflected in the power consumption of the system and can be used to identify programs running.
We test our approach on an HPC rack at Lawrence Berkeley National Laboratory using a variety of scientific benchmarks.
Among other interesting observations, our results show that by monitoring the power consumption of an HPC rack, it is possible to identify if particular programs are running with precision up to 97\% and recall of 95\% even in noisy scenarios.

\input{hpc_intro}
\input{hpc_background}
\input{hpc_setup}
\input{hpc_methodology}

\input{hpc_results}

\input{hpc_conclusion}


\section*{Acknowledgements}

An earlier description of this work was published in Chapter 4 of Bogdan Copos's University of California, Davis Ph.D dissertation~\cite{Copos2017Modeling-System-advisor}.

This work was primarily supported by the Laboratory for Telecommunication Sciences and has been authored by authors at Lawrence Berkeley National Laboratory  under Contract No. DE-AC02-05CH11231 with the U.S. Department of Energy.  
This research used resources of the National Energy Research Scientific Computing Center and was supported in part by the Director, Office of Science, Office of Advanced Scientific Computing Research, of the U.S. Department of Energy under Contract No. DE-AC02- 05CH11231.

Any opinions, findings, conclusions, or recommendations expressed in this material are those of the authors and do not necessarily reflect those of the sponsors  of this work. 

{\footnotesize \bibliographystyle{acm} \bibliography{hpc}}


\end{document}

%% file: hpc_intro.tex
\section{Introduction}

High performance computing (HPC) platforms --- extremely large-scale computation, with an array of specialized components, optimized for large scale-simulation and data analysis --- are indispensable, nation-scale resources.
HPC platforms enable exciting and innovative insight through their ability to process massive amounts of data.
However, these valuable resources come at a high cost.
Aside from the cost of the hardware, there are also high operational costs;
annual energy costs can easily rise to a few million U.S. dollars.
Such high costs elevate the pressure for these platforms to run as efficiently as possible.
These conditions and problems also apply to other large computation facilities such as those offered by cloud computing platforms.

Achieving efficiency is a non-trivial, multi-faceted problem.
One aspect of efficiency focuses on designing new hardware architecture and building new components that consume less power, radiate less heat, and can perform tasks faster.
Efficiency is also considered at the software level.
For example, considerable research efforts are spent on developing efficient job scheduling algorithms that aim for optimal arrangement of program executions.
Previous efforts have also characterized program executions in order to identify bottlenecks and periods of high computational intensity, information that can provide assistance to scheduling or optimization techniques.
An often-forgotten aspect of efficiency focuses on the users.
Misuse and abuse of systems translates to inefficiency.
Consequently, in HPC environments, guaranteeing that users utilize these precious resources as intended (often for scientific purposes) is of high importance~\cite{Peisert2017HPCCACM}.

However, monitoring HPC systems and their users is non-trivial.
Despite the wide range of utilities available, identifying what the users are doing or what programs are running is difficult.
Summary statistics of CPU and memory provided by HPC systems after jobs complete provide useful information but limited insight into the applications running on the HPC system.
These summaries typically include aggregated resource utilization data making it difficult to create unique fingerprints for individual applications.
While time series information about CPU and memory would be equally informative, the collection of this information, especially with high resolution is intrusive and imposes significant overhead on the systems.
``Job logs'' produced by batch scheduling systems can be extremely valuable, but more often than not, such logs only record the application name, which is not always descriptive and useful (``\texttt{a.out}'').

Most institutions that run computational facilities also take procedural measures for controlling access and capabilities of users on HPC platforms.
In most cases, there is a well-defined process for obtaining access.  In closed environments, users are typically highly-vetted employees.   In ``open,'' scientific environments such as those sponsored by the U.S. National Science Foundation and the U.S. Department of Energy's Office of Science, users may be less well known, but a still-rigorous, formal application process is involved.
As part of the application process, users are expected to list the programs they anticipate using for their experiments, and user agreements typically legally constrain the behavior of users.
In order to validate user behavior and whether users are adhering to their agreements, source code can be analyzed statically and binaries analyzed at runtime.  However, static analysis techniques for such analysis are typically significantly incomplete at best, often also requiring runtime analysis as well, and computationally intractable~\cite{metzger2000automatic} at worst.  In addition, runtime analysis techniques impose significant performance overhead.  Both such efforts are challenging, time consuming, and impose significant operational burden.

Despite both technical and procedural measures, user agreement breaches occur.
For example, in 2014, researchers were caught mining electronic currency using a HPC platform owned and operated by the National Science Foundation~\cite{hpcbitmine}.  However, this was neither the first, nor last such incident that has or might occur, given the special capabilities of national high-performance computing resources. It is no surprise, therefore, that ``security'' is a stated goal of the National Strategic Computing Initiative (NSCI)~\cite{NSCI} and has been further examined in Department of Energy Office of Science contexts~\cite{Peisert-et-al.2015ASCR-Cybersecur,Peisert-et-al.2015ASCR-CybersecurityIntegrity}, among others.

\paragraph{Our Work}
In this paper, we present a ``non-intrusive'' method for identification high performance computing platform activity without reducing performance of the HPC system under monitoring.
Our method leverages side-channel information ``leaked'' by the HPC system, specifically, the electrical power consumption of the rack during different computational jobs.
Our approach uses random forest algorithm to learn the power consumption behavior of various programs and uses the learned model to identify jobs whose behavior deviate from those, indicative of unexpected, potentially illicit computational tasks.

\paragraph{Threat Model}
The approach presented in this paper is intended for use by HPC system administrators in order to identify potential misuses of HPC resources.
While previously proposed power analysis methods may yield more accurate results, our method is designed to be non-intrusive and does not require any modifications to the system.
In fact, our approach can be applied to a system in minutes (excluding the training phase).
This paper does not specifically analyze adversarial settings.
Specifically, we do not directly consider scenarios under which a malicious user consciously applies evasive counter measures to avoid detection.  As with any security scenario, such ``masquerade attacks'' are possible, and a variety of techniques exist to counteract such attacks.  However, these methods are outside the scope of this paper.

\paragraph{Evaluation}
We evaluate the effectiveness of our monitoring method by testing it on an HPC platform at the National Energy Research Scientific Computing Center, at Lawrence Berkeley National Laboratory, using a test suite comprised of the NAS Parallel and NERSC Trinity Procurement benchmarks.
We apply our method to various conditions, especially varying levels of ``noise'' (i.e., number of programs/jobs running on the rack).
We determine that most programs can be identified with precision and recall as high as 90\% when only one program is running on the HPC rack.
We also show that programs can also be identified in the presence of noise, with modest precision and recall, depending on the program and the noise level.
For ``noisy'' experiments, we compare our results using two baselines, random guess (lower bound) and mutual information (upper bound) and show that our classification closely follows the theoretic upper bound computed using mutual information.

We note up front that that our technique is different than many traditional intrusion detection papers leveraging anomaly detection in two ways: first, we reiterate that our work to date does not consider adversarial settings --- as with all detection schemes, particularly those based on statistical analyses, there may well be ways for an attacker to ``fool'' this technique into the classifier mis-labeling a program --- eliminating all such opportunity is outside the scope of this paper, which is focused on a novel detection technique in an environment that is not typically studied.
Second, our technique of analyzing a relatively small corpus of known-good programs -- which do not change a great deal over time -- for significant deviations from normal, is also very different than typical applications of anomaly detection to analyzing IP network traffic or sequences of system calls.
Our technique and application is closer in spirit to \emph{specification-based intrusion detection}~\cite{Ko1997}, whereas \emph{anomaly detection} against typical network and host data is more reflective of outlier detection~\cite{Sommer2010} and can suffer considerably from the base-rate fallacy~\cite{Axelsson2000}.

As with many approaches that rely on machine learning, the identification accuracy of our approach depends on the corpus used during training (in terms of both programs and configurations).
For example, compile-time configurations may impact the program behavior, leading to misdetection.
Given the large space of possibilities for training data, it is unrealistic to achieve full coverage.
It is important to note however that most scientific programs have well-defined compilation process, resulting in only a handful of variations.
Furthermore, many high performance computing centers, including the one used in our experiments, have pre-compiled versions of popular scientific applications that are often used by users.

\paragraph{Contributions}

To summarize, this paper's contributions are as follows:
\begin{itemize}
\item A novel non-intrusive method for identifying programs running on an HPC platform
\item A detailed experimental analysis of HPC power usage patterns
\item Results from applying our method to a production HPC rack at Lawrence Berkeley National Laboratory 
\item Comparison of classification results with theoretical upper bound computed using information theoretic metrics 
\end{itemize}

The rest of the paper is organized as follows: 
In Section~\ref{sec:relatedwork}, we give an overview of related previous efforts. 
Section~\ref{sec:setup} explains the components used in our experiments. 
Section~\ref{sec:methodology} discusses the methods used in our experiments. 
Section~\ref{sec:results} presents the results of our approach for identifying programs on HPC platforms. 
Section~\ref{sec:discussion} presents an analysis of the results as well as limitations of the approach presented in this paper. 
Finally, we conclude with Section~\ref{sec:conclusion} where potential future work is discussed. 

%% file: hpc_background.tex
\section{Related Work}
\label{sec:relatedwork}

\subsection{Side Channels}

Side channels have been studied widely.
Previous research efforts primarily exploit side channels to break systems or infringe on the privacy of users.
For example, previous research exploits various side channels, including power, to recover secret cryptographic keys~\cite{kocher1999differential, messerges1999power, luo2014side}.
Other examples include leveraging side channel information like encrypted network traffic to infer users' web browsing activity~\cite{kim2016inferring, cheng1998traffic, bissias2006privacy, hintz2003fingerprinting, liberatore2006inferring, sun2002statistical, herrmann2009website, panchenko2011website}.

While previous work demonstrates that side channels leak information regarding system activity, previous analyses focus on the information flow between the input to a particular system (e.g., cryptographic algorithm, web browser) and the side channel information (e.g., power, encrypted network traffic).
Recently, researchers began exploiting the observation that side channels leak control flow information, and the patterns in side channel information can serve as identifying fingerprints for programs running on a system.
For example, researchers study the use of EM and RF emanations to infer underlying algorithms of programs~\cite{riley2017extraction} as well as deviations in program execution~\cite{nazari2017eddie}.
Our work is similar in that it aims to identify programs executed.
However, our approach is specifically for HPC environments.

\subsection{Power Analysis}

Electrical power was one of the earliest side channels identified.
Kocher et. al.,~\cite{kocher1999differential} introduce differential power analysis and describe specific methods for analyzing power consumption measurements to infer secret keys from cryptographic devices.
Similar to Kocher's work, Carmeli et. al.,~\cite{Carmeli2015On-Bugs-and-Cip} take advantage of bugs in hardware and power consumption of a device to analyze its operations at various stages and retrieve sensitive information.
Power consumption has also been exploited for a variety of other purposes including the identification of Trojans in integrated circuits~\cite{agrawal2007trojan} and exposing a wide spectrum of system-level host information in container clouds~\cite{containerleaks}.

Other efforts study the impact of power analysis side channel attacks from a theoretical point of view.
Micali et. al.,~\cite{micali2004physically}, build a comprehensive but general model for defining and delivering cryptographic security in the presence of side channel attacks.
Standaert et al.,~\cite{standaert2006formal} build upon this work and evaluate the effect of physical leakages with a combination of security and information theoretic measurements.

\subsection{Non-intrusive Load Monitoring}
Power analysis has also been applied to solving problems regarding the power grid.
This area of research became known as \textit{non-intrusive load monitoring} (NILM) and it describes methods for generating ``fingerprints'' for various electric loads in a given household.
One of the earliest published efforts is that of Hart et. al. \cite{hart1992nonintrusive} who apply signal processing techniques on power consumption data of a household to estimate the number and nature of the individual loads within the household.
The authors introduce the \textit{switch continuity principle} which states that ``in a small time interval, [...] only a small number of appliances change state in a typical load'' and build their NILM algorithm based on this principle.
In their work, the authors show that it is possible to determine which appliance is turned on (or off) or in use, given prior knowledge of the appliance models.

Non-intrusive load monitoring shares many similarities with HPC power monitoring.
Instead of measuring the power consumption of households to identify appliances, we monitor HPC rack power consumption in order to determine the what programs are running.
However, it is important to note that there are also some significant differences.
Appliances have a small number of states that, in most cases, last for extended periods of times (e.g., light turned on, refrigerator cooling, dish washer/laundry machine cycle).
For most appliances, the changes in states impose indicative changes in power consumption.
On the other hand, the behavior of programs is considerably more dynamic.
Considering the high operating frequency, modern CPUs can exhibit very short-lasting changes in behavior.

\subsection{High Performance Computing}

Power consumptions and other side channel information, such as I/O behavior, has also been studied for their potential in characterizing the behavior of various applications, especially in HPC settings.
However, the overwhelming majority of these efforts aim to identify phases or bottlenecks in applications and leverage this information to improve the efficiency of HPC platforms.
For example, Isci et. al.~\cite{isci2003identifying} present a method for identifying phases in program power behavior and determining points in the execution of the program that correspond to such phases.
Using their approach, it is possible to generate a \emph{power vector} that represents the estimated power values for 22 processor components such as trace cache and integer execution unit.
This information can be used to dynamically scale the frequency of the processor.

For example, Liu et. al.~\cite{liu2014automatic} present an approach for reliably estimating the user-applications' bandwidth needs, information that can be used to optimize the scheduling of jobs on the HPC platform.
Similarly, Luu et. al.~\cite{luu2015multiplatform} analyze the I/O behavior of applications across multiple runs on an HPC platform and analyze the evolution of applications across time and across platforms.

Our literature search resulted in only a small amount of closely-related work.
In~\cite{6969461}, the authors apply clustering techniques on HPC power data to fingerprint applications.
The authors rely on six time domain features and demonstrate that such features are enough to differentiate between the ten programs used for testing.
However, the authors assume the power traces obtained represent the execution of a single program (i.e., there is no noise or multiple programs executed in parallel).
In our work, we measure the power consumed by an entire rack of HPC nodes and make no assumptions about the number of programs represented in the power trace.
The differences in our work also have implications with respect to feasibility.
Specifically, while their approach requires a single power meter per node, our approach can identify programs running on an entire rack using only one power sensor.
Previous studies indicate that single node jobs compose somewhere between 2\%~\cite{hpc_workload_bluewaters} and 20\% of the workloads~\cite{hpc_workload_nsf}.

Peisert~\cite{Peisert2010-Fingerprint} and Whalen et. al. \cite{whalen2012network, whalen2013multiclass} exploit patterns in Message Passing Interface (MPI) function calls to classify applications by their computation type.
While their effort focuses on characterizing HPC workload, our work aims to identify programs running on an HPC platform at a given time.

%% file: hpc_setup.tex
\section{Experimental Setup}
\label{sec:setup}

Below we discuss the components used in our experiments, with respect to both hardware and software.

\subsection{Sensors}

Power measurements are collected using a micro-phasor measurement unit ($\mu$PMU)~\cite{upmu}.
The $\mu$PMU we use is a device much like PMUs typically used on the transmission grid, but is designed specifically for the distribution grid.  The $\mu$PMUs we use  \emph{sample} electrical current, voltage, and neutral phase angles for all three phases at 512 samples per cycle, 60 cycles per second.
This data is \emph{recorded} as root-mean-squared values at 120 Hertz after phasor conversion.
The sensor is placed in-line with the target device.
In our experiment, the sensor is in-line with a power distribution unit (PDU), which provides power to the HPC rack used in our experiments.

\subsection{HPC Test Platform}

For our experiments, we monitor one of the production compute platform racks at Lawrence Berkeley National Laboratory. 
The rack consists of 36 Condo compute nodes.
Each node is equipped with quad-core Intel Xeon E5530 processors and 24 GB of random access memory.
The nodes are inter-connected using QDR InfiniBand, a well-established computer-networking communications standard for HPC platforms.
However, the nodes we are monitoring do not contain physical storage drives; storage is contained in a separate set of racks.

The power consumption data measured by the $\mu$PMU reflects the work performed by all 36 nodes housed in the monitored rack.
In addition, we wish to highlight that the scheduling system is designed such that at any given time, only one application can run on a particular node.
However, the application  may consist of multiple processes or threads.

\subsection{Test Programs}

In our experiments, we wish to replicate a production HPC environment as much as possible.
To this extent, for our application test suite, we choose benchmarks designed to mimic the behavior of scientific applications.
Specifically, we rely on two benchmark suites: NAS Parallel~\cite{npb} benchmarks and the NERSC-8 Trinity procurement~\cite{trinity} benchmarks.

The NPB are a small set of programs, derived from computational fluid dynamics applications, designed to evaluate the performance of parallel supercomputers.
These benchmarks implement scientific computations, similar to codes that are usually executed by scientists.
Specifically, the benchmarks used in our analysis represent an block tri-diagonal (BT) solver, conjugate gradient (CG) program, embarrassingly parallel (EP) code, fast Fourier transform (FT) kernel, integer sort (IS) program, scalar penta-diagonal (SP) solver, lower-upper Gauss-Seidel (LU) solver, and a multi-grid (MG) application.
The NPB define eight problem size classes (i.e., S, W, A, B, C, D, E, F in increasing order of complexity).
The problem size class defines the input and parameters used during execution and consequently adjust the complexity of the computation.
For example, class S is for small-scale experiments, while class W is for workstation size and class F represents the largest test problems.
In our experiments, we compile the codes of version 3.3.1 for problem size class C and D (for the CG, EP, IS, and MG programs).
We use two problem size classes in attempts to increase the runtime duration of some benchmarks.
For example, one of the benchmarks, the Integer Sort (IS) program runs for less than 10 seconds at problem size class C.
Unfortunately, the runtime duration is only slightly increased at problem size class D.
At the time of this writing, there are twelve benchmarks total, eight of which we use in our analysis.
Four of the benchmarks are eliminated because they either are not available for larger problem sizes or are designed to test I/O performance, which is not visible to our power sensors, since they have visibility only to CPU racks.
Consequently, the NAS parallel benchmarks used in these experiments are: BT, CG, EP, FT, IS, LU, MG, and SP.

The NERSC-8 Trinity Procurement benchmarks are a set of programs created by the National Energy Research Scientific Computing Center (NERSC).
NERSC designed this set of mini-applications to mimic the behavior of real applications and uses these benchmarks for system evaluation and acceptance testing.
We use 4 Trinity programs in our analysis: a parallel algebraic multi-grid solver for linear systems (AMG), 3D Gyro-kinetic Toroidal code (GTC), finite element generation, assembly and solution code (MINIFE), neutral particle transport application (SNAP).

For each program, we gather 50 samples by running each program independently multiple times.
Since these applications are designed as benchmarks, the input is constant between runs.

We note that while the corpus appears small, in many HPC environments, it is usually the case that a small number of applications (between 10-50) compose the majority of the workload (between 60-80\%)~\cite{hpc_workload_bluewaters, hpc_workload_nsf, hpc_workload_nersc}.
While our corpus is composed of benchmark codes, these codes implement common algorithms as characterized by Colella's seven dwarfs and are highly representative of HPC workloads~\cite{hpc_workload_bluewaters}.

%% file: hpc_methodology.tex
\newcommand{\powerfeatures}{24}

\section{Methodology}
\label{sec:methodology}

\begin{figure*}[!ht]
\centering
\includegraphics[width=\textwidth]{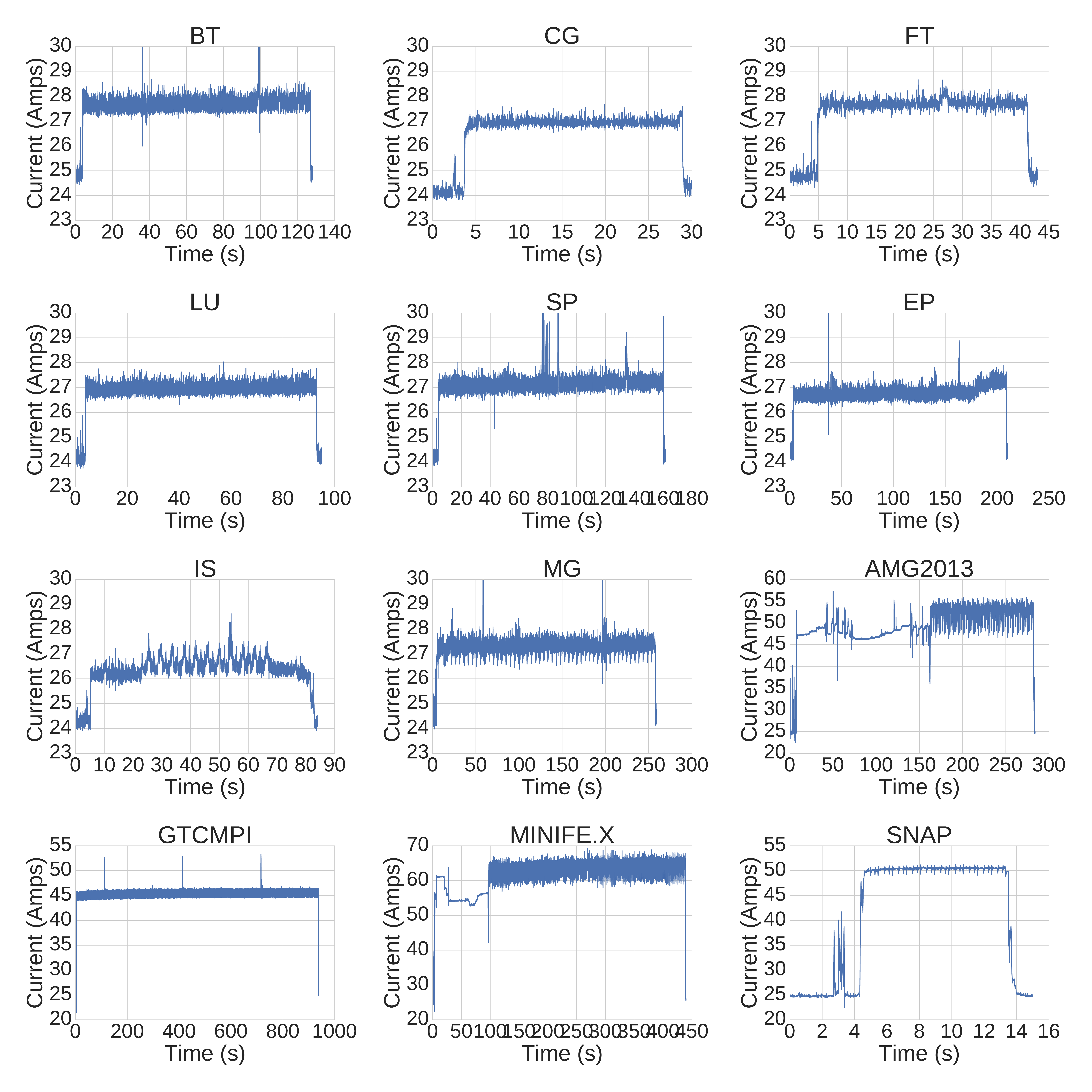}
\caption{Diagram visualizing the raw current magnitude time series for all programs.}
\label{fig:cleansamples}
\end{figure*}

\subsection{Data Collection}

We gather electrical current magnitude time series samples for each program by executing it independently multiple times on our test HPC rack.
In total, we have 50 samples per program.
We refer to these samples as ``clean'' samples since only one program is executed on the monitored rack at a time.
Figure~\ref{fig:cleansamples} depicts raw electrical current magnitude time series samples for all programs.
Several observations can be made from the figure.
First it is important to note that the NAS Parallel benchmarks impose an average consumption of 27 Amps.
On the other hand, the NERSC-8 Trinity Procurement programs impose a much higher load which causes the system to draw more current.
Furthermore, we can see that while some programs exhibit (visually) distinct patterns, other programs such as CG and EP do not and are relatively similar.

In production HPC environments, it is rare that a single job is executing on any given rack.
To replicate such ``noisy'' environments, we combine ``clean'' program samples to synthetically generate ``noisy'' samples.
In particular, to generate a ``noisy'' sample for program $P_{target}$, we first select a sample for program $P_{target}$ that will serve as the \emph{target sample}.
Noise is added by randomly selecting a ``noise program'' $P_{noise}$ from the pool of available programs.
We make sure the ``noise program'' $P_{noise}$ is not the same class type as the target sample.
Next, we randomly select a ``noise program sample'' and add it to the target sample, starting at a randomly selected index.
Randomness is introduced to replicate the unpredictable changes in the work load of an HPC rack.

We choose to generate synthetic noisy samples as opposed to actually running multiple programs in parallel for two reasons.
First it allows us to generate significantly more combinations of samples faster and a wider variety of ``noisy'' conditions.
Additionally, it avoids spending costly cycles on the HPC platform.
We note that this technique was designed from the outset to be realistic by working in partnership with operational staff at Lawrence Berkeley National Laboratory.  Those staff confirm the validity of the approach assuming the number and type of components (compute nodes, network routers, service nodes, etc...) for any given rack being monitored is roughly the same, which is the case for typical HPC centers.

The synthetic noisy samples generated are equivalent to a sample obtained by actually running the program concurrently.
Any new job is assigned to one or more idle node(s).
Given that electrical power of individual circuits is additive (regardless of configuration), the sum of the power of individual circuits is equal to the power of the combined circuits.
In other words, the sum of the power consumption of individual nodes is equal the power consumption of multiple nodes.
Previous work in the area of non-intrusive load monitoring uses similar techniques~\cite{nandy,zia2011hidden}. 

\subsection{Feature Selection}

To facilitate the generation of ``fingerprints'' or each program, we first extract a variety of features to describe the samples.
Specifically, we rely on both time and frequency domain features.
Time domain features capture temporal information that is not captured by frequency analysis.
However, unlike time domain features, frequency domain features are less impacted by noise.T
Therefore, we rely on both time and frequency domain features to capture as much information and to better distinguish between signals.

Selecting time domain features is non-trivial.
The well-known Anscombe's quartet elegantly describes the challenge.
Specifically, Anscombe's quartet presents four data sets that share statistical metrics, some of which are identical (i.e., mean, variance) while others are extremely close.
In our problem setting, noise refers to the fact that the amount of current consumption, as measured, is a holistic representation of the system's activity.
In other words, the current magnitude time series reflects all of the jobs running across all nodes of the monitored rack.
To establish a representative set of time domain features, we compare their distribution across all samples from all of the programs.
The following features represent the final set of time domain features: maximum value, percentiles, auto-correlation coefficients with different lag values, standard deviation, empirical cumulative distribution function, stationary coefficient, discriminant function analysis, absolute energy, and ratio of variance to standard deviation. 

For feature extraction, we focus on three types of frequency analysis: discrete Fourier transform, power spectral density, and wavelet analysis.
In comparison with discrete Fourier transformation, power spectral density estimates the power of various frequency components of a signal.
To compute power spectral density, we apply Welch's method~\cite{welch1967use} to each sample.
We use the implementation found in the SciPy~\cite{scipy} Python library (version 0.18.1).
Welch's method is used to estimate the power of a signal at different frequencies and it does this by performing windowed short term discrete Fourier transform.
Instead of using the raw power spectral density values, we extract the top four peak values (both frequency and corresponding power spectral density).

Fourier transform intrinsically compromises time and frequency resolutions and is restricted to decomposing a signal into sinusoids.
To combat these limitations, we also perform wavelet analysis.
To perform wavelet transformation, we leverage the Ricker wavelet function.
We use the implementation found in the SciPy~\cite{scipy} Python library (version 0.18.1).
For our features, we extract coefficients with different width values for the wavelet function.
Ultimately, for our classification, we rely solely on wavelet coefficients.

As mentioned above, feature selection is a meticulous process that requires a lot of experimenting and finesse.
The final features are selected using two types of analysis.
First, we perform distribution and variance analysis of the feature values across the different samples in order to identify and remove features with values shared by many types of programs.
We also use Gini impurity to determine how each feature contributes to the labeling of a sample.
Gini impurity is a method used during the building of decision trees to measure the prediction power of the features.
Gini impurity aims to measure the homogeneity of features and it does this by estimating how likely it is to mis-label a sample if the label is assigned randomly according to the distribution of the labels determined by the feature.
When a feature is able to narrow down the label to only one option, its Gini index is 0.
When there are multiple labels and the probability associated with each label is the same, the index is one.
Therefore, features with low index values are more valuable in classifying data.

After thorough examination of the features and several experiments, we converged on a set of \powerfeatures{} total feature values.\footnote{some features (e.g., percentiles, wavelet coefficients) have multiple feature values}

\subsection{Analysis}

There are several ways to process the samples.
One method is to extract features for each sample and assign the corresponding label.
However, such strategy can have several negative effects.
As described above, considering entire samples when computing certain features can result in information loss (refer to the mention of Anscombe's quartet above).
In addition, the results of certain frequency domain transformations depend on the length of the sample and are impacted when the samples vary in duration.
In our experiments, the runtime duration of programs varies not only across programs but even for the same program.

In addition, we believe the Hart's \emph{switch continuity principle} also applies to this setting.
Specifically, when considering short windows, only a small number of programs are expected to exhibit significant changes.
Consequently, our method divides each sample into equal length windows.
Each window is preprocessed (e.g., normalized) after which features are extracted.
To introduce time-dependency information, we group consecutive windows together into ``window groups.''
Each ``window group'' represents a single feature vector (used in classification).
In other words, a feature vector is composed of sets of feature values, one set per window.

Relying on windows also has other benefits, specifically with respect to the program input.
The behavior of computational programs is usually split between three phases: input reading (denoted as $I$), processing (denoted as $P$), and output writing (denoted as $O$).
While the size and type of input may vary, the behavior during the three phases is similar.
If the window size is carefully selected, breaking the power signal sample into windows means that our approach makes no assumption about duration (or even number or order of windows) for each of the three phase for a given program run.
Using the denoation from above, our approach can successfully identify executions of program $X$ whether the behavior is represented by $IIIPPPOOO$ or $IPO$ or even $IPOIPOIP$.

We experimented with several window sizes, window group sizes, as well as various overlap strategies (i.e., where window groups share overlapping windows).
Our experiments show that classification is most accurate when using 2 second windows (i.e., 240 data points at 120 Hz) and window groups of size 4.
However, the variations in classification results due to window sizes and window group sizes are relatively small.
We observe no significant improvement when overlapping window groups and consequently choose to not implement any such strategy.

\subsection{Building Program Fingerprints}

We apply machine learning to generate fingerprints for each program using the feature vectors constructed as described above.
Specifically, we rely on a random forest algorithm.
A random forest is an ensemble learning algorithm often used for multi-class classification.
The random forests are constructed during the training phase.
Specifically, to grow a random forest, subsets of features are selected (with replacement) and decisions trees are created for each feature subset.
The label of the target sample is determined by the mode of the decision tree predictions.
Unlike decision trees, which are very susceptible to over-fitting, random forests reduce the bias by using a bagging process.
Bagging occurs at two levels: data selection and variable selection.
Using bagging, the construction of each tree uses a different subset of data and as well as a different subset of variables.
For our experiments, we use the \textit{scikit-learn}~\cite{scikit-learn} (version 0.17.1) Python machine learning library for both the random forest implementation as well as supporting tools.

%% file: hpc_results.tex
\section{Results}
\label{sec:results}

In this section, we present the results of our experiments.  As mentioned earlier, we reiterate that neither our approach nor our target environment, and therefore also our results, should be compared directly with typical applications of anomaly detection, as our technique of analyzing a relatively small corpus of known-good programs that do not change a great deal over time is very different than common applications of anomaly detection, which are more reflective of outlier detection, and suffer considerably from the base-rate fallacy.

We present the results of identifying programs in both clean (i.e., $ noise = 0 $, target program is the only program running) and noisy experiments (i.e., $ noise \geq 1 $).
In total, we run 8 trials for each experiment (i.e., noise level) and each experiment performs 3-fold cross validation.
The results of the 3-fold cross validations are verified for statistical significance using the Wilcoxon-Mann-Whitney (WMW) signed ranked test ($p < 0.01$) for the clean classification experiments.
Unlike the $t$-test, the Wilcoxon signed rank test does not assume that the data are sampled from a Gaussian distribution.

Some programs were excluded from the experiments because the samples are too short in length.
For example, the execution for the SNAP program takes only twelve seconds.
Such short runtime duration is not enough to compose even two feature vectors of four windows of two seconds each.
We filter out samples (consequently programs) with length shorter than 32 seconds.
However, we wish to point out that the window size and window group size can be adjusted as seen fit.

\subsection{Clean classification}

\begin{table}[!ht]
\centering
\label{tab:cleanresults}
\begin{tabular}{|c|c|c|c|}
\hline
  \textbf{Program} & \textbf{Precision (\%)} & \textbf{Recall (\%)} & \textbf{F Score} \\ \hline
  BT & 93.55  & 97.04 & 0.95 \\ \hline 
  CG & 50.0  & 3.32 & 0.06 \\ \hline 
  EP & 51.11  & 76.17  & 0.50 \\ \hline
  FT & -- & --  & -- \\ \hline
  IS & 17.04  & 100.0 & 0.28 \\ \hline 
  LU & 89.18 & 100.0 & 0.94 \\ \hline
  MG & 100.0 & 54.50 & 0.70 \\ \hline
  SP & 94.30 & 84.03 & 0.89 \\ \hline
  AMG & 100.0 & 100.0 & 1.0 \\ \hline
  GTC & 100.0 & 100.0 & 1.0 \\ \hline
  MINIFE & 100.0 & 100.0 & 1.0 \\ \hline
  SNAP & -- & -- & -- \\ \hline
\end{tabular}
\caption{Our approach achieves high precision and recall for most programs. Manual analysis of the results show that BT, CG, and EP are often misclassified as each other.}
\end{table}

\begin{figure*}
\centering
\includegraphics[width=\textwidth]{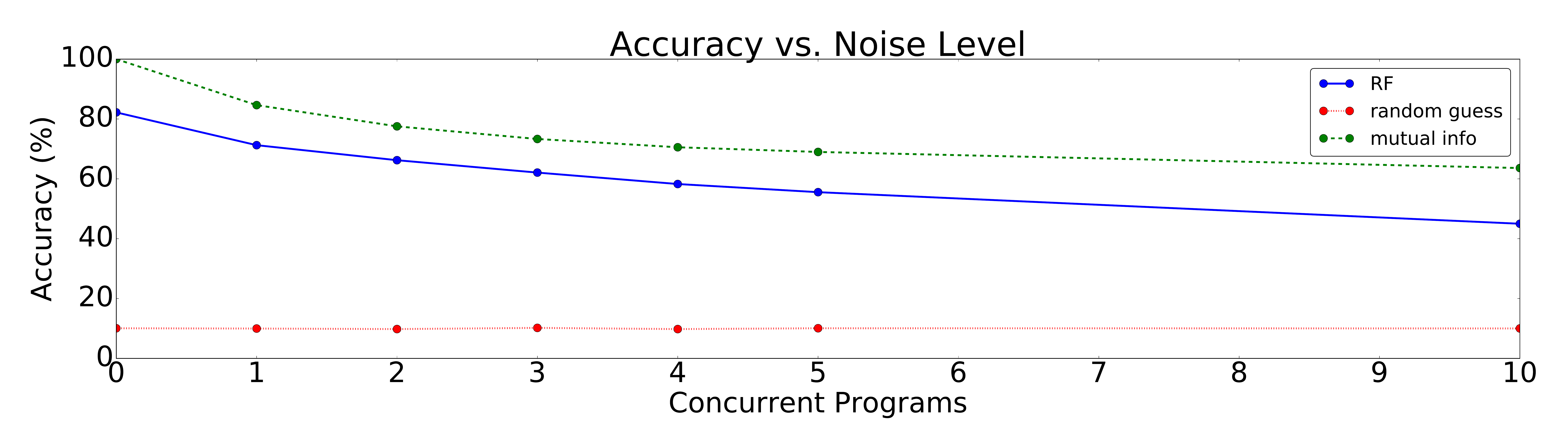}
\caption{While noise (i.e., multiple concurrent jobs) does impact the average classification accuracy of our approach, the impact mirrors the loss in the theoretical mutual information.}
\label{fig:accuracyvsnoise}
\end{figure*}

Table~\ref{tab:cleanresults} describes the results of the clean experiments.
Specifically, the classification precision, recall, and F-beta scores are listed for each program.
Precision, or \emph{true positive rate}, is defined by: $$ precision = \frac{true\_positives}{(true\_positives + false\_positives)}$$
Precision describes the ability of the classifier to only identify samples that belong to the target sample.
Recall, also known as \emph{sensitivity}, is defined by: $$ recall = \frac{true\_positives}{(true\_positives + false\_negatives)} $$
Recall represents the fraction of relevant entities that have been identified over the total number of relevant samples.
F-beta score (F-score for short) is a value between 0 and 1 that represents the harmonic mean between precision and recall.
In our case, precision and recall are weighted equally in the calculation of the F-score. 

Our method is capable of identifying many of the programs with high precision and recall.
6 of the 10 programs are identified with an F score of over 85\%.
However, the classification of three programs (i.e., CG, EP, and IS) is considerably less accurate.

\subsection{Noisy classification}

As previously mentioned, in most production HPC environments, it is rarely the case that only one program (or job) is running on a given rack.
Consequently, we test our approach on multiple noise levels.
The noise level represents the number of samples (of other classes/programs) used to generate the noisy synthetic sample.
For example, a noise level of one means that the target program sample was mixed with one other sample of a randomly selected sample.
It is important to note that the class type of the sample used as ``noise'' is different from the target sample class type.

Figure~\ref{fig:accuracyvsnoise} visualizes the impact of noise on the average accuracy of our classification.
We present two comparison baselines for our results.

\begin{figure*}[!ht]
\centering
\includegraphics[width=\textwidth]{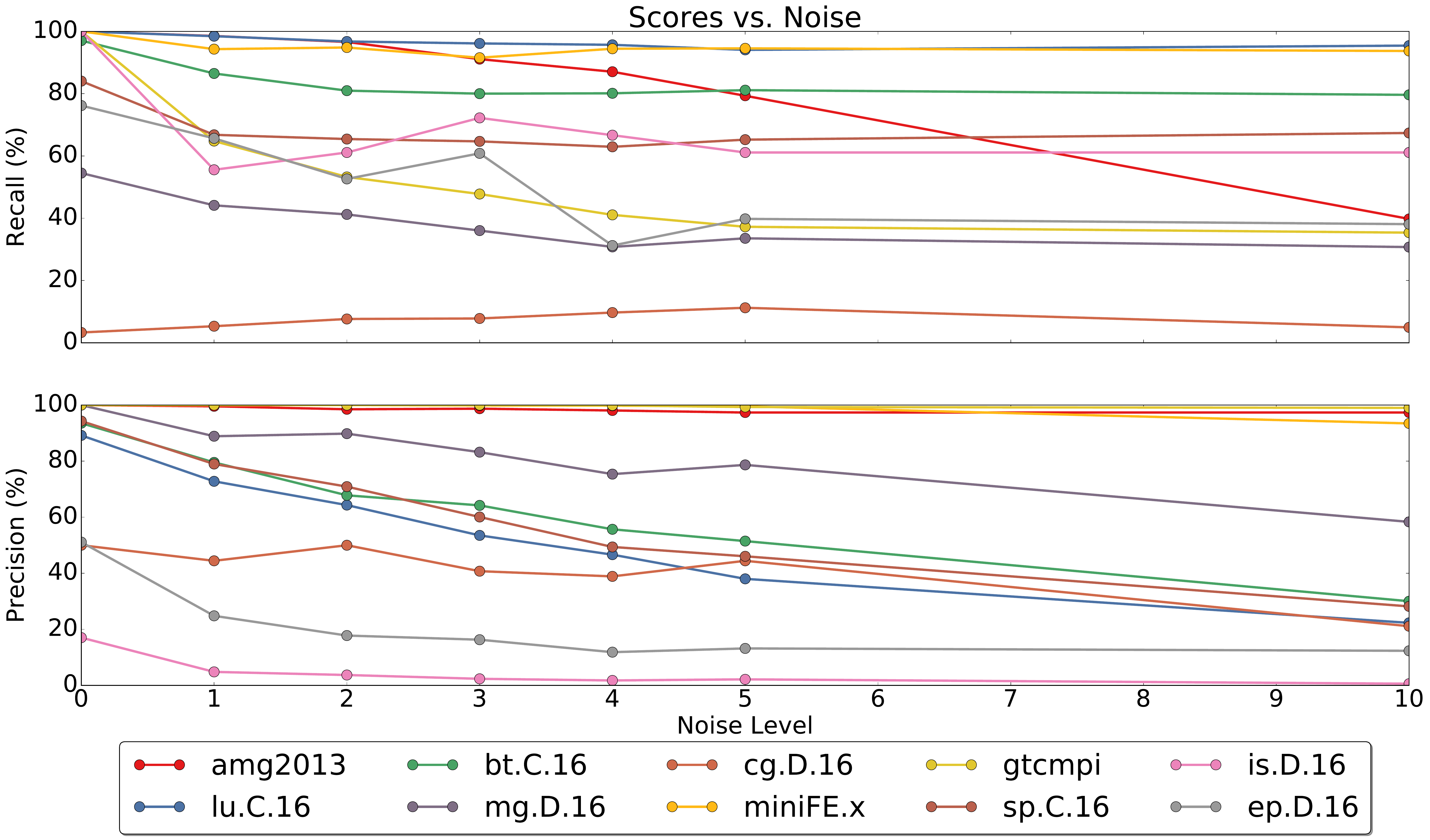}
\caption{Noise impacts each program differently depending on (i) the nature of the program and its behavior as exhibited in the power consumption and (ii) the nature of the noise.}
\label{fig:programsvsnoise}
\end{figure*}
As a theoretical lower bound, we compare the classification results with a random guess strategy, where the label of each window is randomly selected from the pool of available programs.
Given the randomly guessed window labels, the sample label is again selected by picking the label with the highest count.
The accuracy of the random guess strategy is visualized in a green dotted line.
The expected value of the average accuracy of randomly guessing is:  $$ E[rand\_guess] = \frac{1}{num\_programs} $$
In our case this is equal to $\frac{1}{8}$.
Naturally, the accuracy of random guessing is unaffected by noise.

In contrast, we estimate the theoretical upper limit of such classification using \emph{mutual information}, represented by a green dashed line.
The \emph{mutual information} of two samples is an information theoretic measure of the mutual dependence between the two samples.
Specifically, mutual information computes the conditional Shannon entropy of one sample based on the second sample.
In our work, we use the mutual information as a way to estimate the information lost by the addition of noise to a sample.
As such, for each sample used for classification, we compute the mutual information between the original sample and the corresponding synthetic noisy sample.
As expected, in our experiments, the mutual information varies depending on the type of sample and the placement, as well as magnitude, of noise.

As shown in Figure~\ref{fig:accuracyvsnoise}, we can see that our classification approach performs better than the random guessing strategy and is within 15-20\% of the theoretical upper bound defined by the mutual information metric.
It is important to note that some of the difference between the upper bound and our classification results is accounted by the training error.
In other words, two executions of the same program do not generate perfectly identical samples.
Consequently, the mutual information between such two samples is often less than one.
This is also reflected in the clean classification results described in Table~\ref{tab:cleanresults}.

Figure~\ref{fig:programsvsnoise} depicts the precision and recall results for each individual program across varying levels of noise.
It is important to note that while both precision and recall decrease for all programs as the level of noise increases, the rate of decline varies across the programs.
For example, several programs, such as MG and MINIFE, maintain recall scores above 90\% despite the increase in noise.
Moreover, programs AMG, MINIFE, and AMG can be identified with precision above 90\% despite the level of noise.
Noise greatly impacts the identification (both precision and recall) of other programs, such as SP.

In the case of some programs, the performance varies between two consecutive noise levels.
For example, in the top part of Figure~\ref{fig:programsvsnoise}, we can see that the recall of the MG program increases from noise level 2 to 3 and again between noise levels 4 and 5.
This behavior is explained by the randomness involved in the generation of noisy synthetic samples.
Specifically, our approach randomly selects the starting location at which noise is added to the target sample.
This can have significant impacts on classification.
In other words, if noise is added from the beginning of the target sample, more of the window sets are affected and consequently the overall accuracy decreases.
On the other hand, if noise is added only in the last few window sets, it is possible that the previous window sets are correctly identified and the only misclassified window sets are those affected by the noise.

%% file: hpc_conclusion.tex
\section{Discussion}
\label{sec:discussion}

As the results show (see Figure~\ref{fig:programsvsnoise}), our method performs well for most programs in noiseless scenarios (i.e., when the target program is the only program executing on the HPC rack).
However, this is not the case for all programs.
We analyze the results manually to determine why some of the programs have low scores even in noiseless scenarios.
Our analysis of the confusion matrix shows that programs EP and CG are often misclassified as each other.
In addition, we see that program IS is identified with high recall but low precision, signifying that there is a large number of false positives.
This is also represented in the confusion matrix, which shows that window sets of \emph{most} other programs are misclassified as IS.

On the other hand, the average accuracy of our approach decreases as the number of programs executed increases.
However, it is important to note that the impact of noise differs across the different programs.
For example, even with 10 other programs running simultaneously, the MINIFE program can still be identified with relatively high precision and recall (93\%).
This is explained by the fact that programs such as AMG, GTCMPI, and MINIFE utilize more resources and consequently exhibit more prominent and dominant patterns in their electrical current footprint.
This is also visibly apparent in Figure~\ref{fig:cleansamples}.
In other words, our method is best at identifying programs that utilize significant resources (e.g., number of nodes, CPU cycles).

Another interesting observation is that noise impacts precision more so than recall.
In other words, our approach produces few false negatives and a higher number of false positives.
Depending on the application, this may be a desirable effect.

Overall, from the results, it is also evident that our approach is more successful in identifying programs that exhibit distinctive patterns in their behavior.
For example, even amongst the less CPU intensive NPB programs, the BT program is more accurately identified, which we believe is explained by the frequency domain pattern the sample exhibits.
On the other hand, our approach struggles with programs such as EP for which the current magnitude time series is relatively constant and show no patterns.
As previously mentioned, this is also confirmed by the confusion matrix, specifically the distribution of false positives.

These observations have several implications when applying our approach to a real HPC environment.
First, as described above, our approach is more successful at identifying resource-intensive applications.
Consequently, the method described in this paper may not be able to identify executions of short programs or scripts.  That being said, operators of both high-performance computing centers and commercial cloud providers have expressed their primary concern as mis-use of computation resources for unauthorized purposes.  Therefore, it is the computationally-intensive tasks, which our technique performs better on, that is more valuable to potential end users.
We also note that most HPC applications are highly-parallelizable computationaly intensive applications that often require several nodes to run efficiently, and so leveraging our technique to identify tasks that are not using the CPU efficiently is a potentially interesting application.
Finally, we also believe in our current approach the granularity of the power data is a limiting factor for improved accuracy and that higher-resolution (more than 120Hz) power data could increase the accuracy of our approach and broaden the coverage of our approach to also include programs with short execution times or low resource usage.

\section{Limitations and Future Work}

It is important to note that there are alternative methods for measuring the system power consumption.
For example, it is possible to insert probes into the CPU and measure the power directly.
While such alternatives provide a clearer picture of the CPU activity and may produce better results, they are also intrusive and less practical.
Introducing probes into the CPUs of every single compute node in every rack is labor-intensive and increases the overhead of the system maintenance process.

Aside from data collection alternatives, our approach has a few limitations. 
Our approach is limited with respect to identifying HPC activity in noisy scenarios.
Throughout our experiments, we have applied various noisy filtering techniques, in both time and frequency domain, with limited success.
Future research is needed to identify compatible noise filtering and power disaggregation techniques.

Another limitation of our approach lies in the dependency on supervised learning.
Specifically, our approach can only identify the programs the algorithm was trained on.
As discussed earlier, we do not believe this to be a critical limitation of our approach considering the homogeneous nature of most HPC workloads (i.e., few applications consume majority of cycles~\cite{hpc_workload_bluewaters,hpc_workload_nsf,hpc_workload_nersc}).

In the future, we wish to address some of the limitations and improve our approach.
First, we are interested in studying how changes in the compilation settings (e.g., optimization flags), source code (e.g., different versions), and input affect the behavior of the applications as exhibited in the power consumption.
Furthermore, we would like to expand our work to classify HPC activity into computational classes (similar to~\cite{whalen2012network, whalen2013multiclass,Peisert2010-Fingerprint}), rather than individual programs.
We believe such an approach would identify what types of applications users are running on a particular system without explicitly analyzing and extracting ``fingerprints'' for every possible application.
Additionally, by classifying computational classes rather than individual programs, our approach would be more robust to changes in the source code, compilation settings, or even input.
Additionally, we believe that higher resolution data could disclose additional patterns depicting the behavior of the programs that could be leveraged to further differentiate between workloads and we would like to test our hypothesis with various power sensors.
Other future work includes levearging other side channel information (e.g., performance hardware counters), combining mutliple side channels to further improve classification, as well as applying other machine learning techniques, such as deep learning.

Finally, we reiterate that our work to date does not consider adversarial settings.

\section{Conclusion}
\label{sec:conclusion}

High performance computing facilities are national resources, and their availability for legitimate use is limited.  Past incidents have provided evidence that high-performance computing resources are sometimes abused by users for personal gain.
However, identifying how those systems are used, by understanding what programs are running on such systems, is non-trivial and many monitoring utilities provide limited, if any, insight.

In this paper we present a method for inferring the activity of HPC users by non-intrusively monitoring the power consumption of HPC systems and leveraging that information to identify what programs are running.
At its core, our method leverages the observation that programs exhibit patterns in CPU and I/O throughout execution and these patterns are reflected in the power consumption of the HPC node, specifically in the amount of electrical current drawn by the system.

Specifically, our approach measures the electrical current drawn by a HPC rack (of multiple compute nodes) during the execution of programs and applies machine learning to generate ``fingerprints'' for each program and later identify programs given unknown electrical current samples.

Our experiments show that the method presented in this paper is capable of identifying programs with high precision (97\%) and recall (95\%) even in noisy scenarios, where multiple programs are running simultaneously on the same rack.
While the average accuracy decreases as the number of simultaneously running programs increases, our method is capable of identifying large, resource-intensive programs with high precision and recall.
Although the experiments presented in this paper focus on HPC environments, we believe similar approach can also be applied in cloud computing platforms, for example to identify when users leverage cloud computing resources for nefarious purposes such as running a botnet.

At a broader level, our method demonstrates that side channel information can be leveraged to accurately determine what a system is doing.
We envision our approach being used in several ways.
As presented in this paper, we believe our approach could be used by HPC/cloud system administrators as a first-pass filter towards identification of potential misuse of resources.
This would involve training the model on the common applications and using it to fingerprint on-going workloads.
If a particular workload is not identified as any of the trained common applications by our approach, an alert is raised and manual analysis of the particular code can be performed.
Another benefit of our approach is that it provides an alternative method for workload characterization that can be applied as the systems are running.
Specifically, information regarding type and duration of workloads can be valuable for optimizing the HPC systems, particularly with scheduling.